\documentclass[fleqn,11pt]{wlscirep}
\usepackage[T1]{fontenc}
\usepackage{graphicx}
\usepackage{hyperref}
\hypersetup{colorlinks,breaklinks,
            urlcolor=[rgb]{0.,0.4,0.4},
            linkcolor=blue,
            citecolor=red}
\usepackage{charter}

\title{Exactly solvable model of two interacting Rydberg-dressed atoms confined in a two-dimensional harmonic trap}

\author[1]{Przemys\l aw Ko\'scik}
\author[2,*]{Tomasz Sowi\'nski}
\affil[1]{Institute of Physics, Jan Kochanowski University, ul. \'Swi\c{e}tokrzyska 15, PL-25406 Kielce, Poland}
\affil[2]{Institute of Physics, Polish Academy of Sciences, Aleja Lotnik\'ow 32/46, PL-02668 Warsaw, Poland}
\affil[*]{tomsow@ifpan.edu.pl}

\begin{abstract} 
Exactly solvable model of two Rydberg-dressed atoms moving in a quasi-two-dimensional harmonic trap is introduced and its properties are investigated. Depending on the strength of inter-particle interactions and the critical range of the potential, the two-particle eigenstates are classified with respect to the excitations of the center-of-mass motion, relative angular momentum, and relative distance variable. Having these solutions in hand, we discuss inter-particle correlations as functions of interaction parameters. We also present a straightforward prescription of how to generalize obtained solutions to higher dimensions.
\end{abstract}
\begin{document}

\flushbottom
\maketitle
\thispagestyle{empty} 

\section{Introduction}
Few-body systems of ultra-cold atoms provide a very comprehensive toolbox for exploring fundamental properties of quantum systems containing a mesoscopic number of particles \cite{2012BlumeRPP,2016ZinnerRev,2019SowinskiARX}. Due to accessible tunability of their different parameters they can serve as quantum simulators of strongly correlated quantum systems of a few particles described by models being far beyond computational facilities of computers nowadays  \cite{1982FeynmanIJTP}. Typically, in the context of ultra-cold physics one assumes that mutual interactions between atoms are dominated by short-range forces and may be represented by simple $s$-wave (for bosons) or $p$-wave (for fermions) scattering processes. However, mainly due to the experimental progress with polar atoms and molecules, also dipolar long-range and anisotropic interactions are widely considered and they were found to have very interesting consequences for the system's properties \cite{2008BaranovPhysRep,2009LahayeRPP,2016GadwayJPhysB}. Alternatively, long-range interactions between ultra-cold atoms can be achieved when their excitations to the Rydberg states are considered, {\it i.e.}, when atoms become excited to large principal numbers \cite{2005GallagherBook}. In such a case, mutual interactions are not only long-range and strong but also have multipolar properties. In fact, the resulting interaction potential can be viewed as a combination of the standard long-range van der Waals interaction acting over large distances (measured recently with a very sensitive experimental scheme \cite{2013BeguinPRL}) and soft-core, almost constant, potential when atoms are close enough \cite{2004TongPRL,2010HonerPRL,2010PupilloPRL,2010HenkelPRL,2012LiPRA}. Since coherent excitations to Rydberg states on demand were recently announced by many experimental groups (for review see \cite{2010SaffmanRevModPhys}) Rydberg atoms become one of the candidates for fundamental blocks of future quantum simulations. In such cases, a deep understanding of their spatial correlations may be fundamentally important.

Inspired with the above experimental motivations, in this work we study the problem of two interacting Rydberg atoms confined in a two-dimensional isotropic parabolic trap. The discussion is carried out in the framework of the simplified but exactly solvable model. The simplification is based on the assumption that the main contributions to the spatial properties of the system come from the soft-core part of interactions. This part we model simply by a flat potential of a finite range and strength. At the same time, we assume that the long-range part of the interactions is adequately less important and can be safely omitted. In this way, we end up with the interaction potential modeled by a step function in the relative distance between particles. It turns out that assuming this simplified shape of inter-particle interactions one can fully solve the corresponding two-particle Schr\"odinger equation in terms of special functions. Along with the discussion, we argue that the simplified approach is justified provided that the critical radius is not very large when compared to the natural length of the trapping potential.

Our work is organized as follows. In Sec.~\ref{Sec:Model} we introduce the simplified model of interacting Rydberg-dressed atoms confined in a two-dimensional parabolic trap and we explain its origin. In Sec.~\ref{Sec:Eigen} we present a full solution of the corresponding two-particle eigenproblem in terms of the hypergeometric confluent functions. Then in Sec.~\ref{Sec:Porownanie} we perform classification and discussion of the spectrum of the system and its lowest eigenstates. At this point, we also validate our simplified model by comparing its predictions with numerical results obtained for the same system but with interactions modeled by the realistic potential. In Sec.~\ref{Sec:Cor} we focus on inter-particle correlations for different parameters of interactions, while in Sec.~\ref{Sec:Gen} we shortly explain how one can generalize our analytical solutions to the case of two Rydberg-dressed atoms in a three-dimensional harmonic trap. Finally, we conclude in Sec.~\ref{Sec:Conclu}.

\section{The model} \label{Sec:Model}
In this work we consider the system of two interacting quantum particles of mass $m$ confined in an external quasi-two-dimensional harmonic isotropic trap of frequency $\Omega$. The Hamiltonian of the system reads
\begin{equation} \label{Hamiltonian}
\hat{\cal H} = \sum_{i=1}^{2}\left(-\frac{\hbar^2}{2m}\nabla_i^2 + \frac{m\Omega^2}{2}\boldsymbol{r}_i^2\right) + {\cal V}\left(|\boldsymbol{r}_1-\boldsymbol{r}_2|\right),
\end{equation}
where $\boldsymbol{r}_i=(x_i,y_i)$ are positions of particles. We assume that the dimensional reduction of the problem to two spatial dimensions is granted by very deep confinement in the remaining third spatial dimension. In such a case any spatial excitations in this direction are strongly suppressed. Consequently, one can safely assume that the dynamics is frozen and particles occupy only the lowest single-particle orbital.

It is well-known that in the case of highly-excited ultra-cold off-resonantly dressed Rydberg atoms the effective interaction potential have a very characteristic form \cite{2004TongPRL,2010HonerPRL,2010PupilloPRL,2010HenkelPRL,2012LiPRA,2017PlodzienPRA}. It can be viewed as the natural van der Waals interaction acting on large distances ($\sim 1/r^6$) with significant modification when the inter-particle distance is comparable with so-called critical Rydberg radius $R_c$. It is argued that the potential can be written as:
\begin{equation} \label{RydbergPot}
{\cal V}_{R}(r)=\frac{g}{1+ \left(r/R_c\right)^6}.
\end{equation}
Here, the two independent parameters $g$ and $R_c$ describe characteristic scales of the potential and they are related to the interaction strength and aforementioned critical distance at which interactions change their character. Since these parameters directly depend on experimentally accessible quantities, namely the effective Rabi frequency $\omega_R$ and the detuning $\Delta$, as $g = 2\omega_R^4/\Delta^3$ and $R_c = (C_6/2\Delta)^{1/6}$ ($C_6$ is the dispersion coefficient being constant for chosen atom), they can be treated as parameters which can be tuned on demand (for details see for example \cite{2012LiPRA}).

\begin{figure}[t]
\includegraphics[scale=0.9]{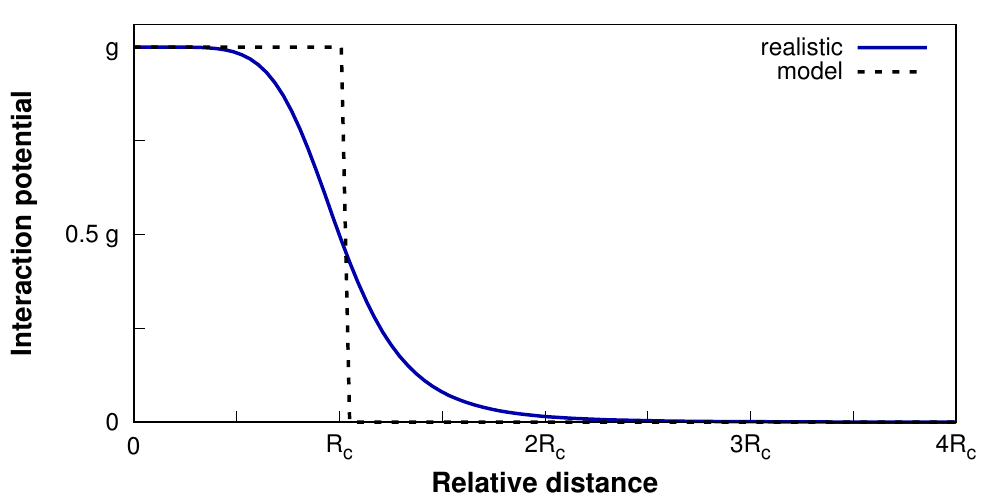}
\caption{Schematic comparison of the realistic shape of interaction potential ${\cal V}_R(r)$ between Rydberg-dressed atoms \eqref{RydbergPot} and the simplified rectangular box potential ${\cal V}(r)$ given by \eqref{interaction}. In the latter case, the long-range part is completely neglected while the short-range part is replaced by a constant energy shift. Substantial differences between the two models are visible only in the vicinity of the critical radius $R_c$. \label{_Fig1}}
\end{figure}

In fact, the potential \eqref{RydbergPot} belongs to the large class of potentials of the form ${\cal V}^{(\alpha)}_R=g[1+(r/R_c)^\alpha]^{-1}$ having an almost flat soft-core in the center  ($r\lessapprox R_c$) and a long-range tail decaying algebraically as $\sim\!r^{-\alpha}$. In the limit of very large powers $\alpha$ this class of potentials is exactly equivalent to the simplified potential of the form
\begin{equation} \label{interaction}
{\cal V}(r) = \left\{
\begin{tabular}{ll}
$V$, & $r<a$ \\
$0$, & $r\geq a$
\end{tabular}
\right.
\end{equation}
provided that $V=g$ and $a=\pi R_c/[\alpha\,\mathrm{sin}(\pi/\alpha)]$. It means that in the case studied ($\alpha=6$) the potential \eqref{RydbergPot} can be modeled by approximate potential \eqref{interaction} by fixing $a=\pi R_c/3$ (see Fig.~\ref{_Fig1}). Since the power $\alpha$ is quite large the approximation can be treated as reasonable.

Having this argumentation in mind, in our work we model mutual interactions between particles with the finite-range soft-core potential \eqref{interaction}.
In this approximation, the interaction energy vanishes whenever the distance between particles is larger than the potential range $a$ and it has non-vanishing constant value $g$ at short distances. In the following, we show how to find {\it all eigenstates} of the Hamiltonian \eqref{Hamiltonian} with corresponding eigenenergies. In this way, we generalize recent results obtained for infinite interaction strength ($V\rightarrow\infty$) \cite{2019SaraidarisARX}, as well as for the corresponding one-dimensional problem \cite{DeuretzbacherPhD,2018KoscikSciRep}, and we extend the list of exactly (or almost exactly) solvable models of two interacting particles \cite{1998BuschFoundPhys,1998BaoPRA,1999BaoPRA,2001GirardeauPRA,2005IdziaszekPRA,2006IdziaszekPRA,2011ChenEPJD,2016JieJPhysB,2016OldziejewskiEPL,2018DawidPRA,2019BudewigMolPhys}. We also perform a comprehensive comparison of the results obtained with the two potential shapes \eqref{RydbergPot} and \eqref{interaction}.

\section{Eigenproblem} \label{Sec:Eigen}
The first step towards diagonalization of the Hamiltonian \eqref{Hamiltonian} is to separate the center-of-mass motion. Since particles are confined in a harmonic trap the separation is done by performing the following transformation of coordinates to the center-of-mass and relative motion positions
\begin{subequations}
\begin{align}
\boldsymbol{R} &= \frac{\boldsymbol{r}_1+\boldsymbol{r}_2}{2}, \\
\boldsymbol{\xi} &= {\boldsymbol{r}_1-\boldsymbol{r}_2}.
\end{align}
\end{subequations}
Indeed, in this coordinates the Hamiltonian separates into two independent parts $\hat{\cal H}=\hat{\cal H}_{R}+\hat{\cal H}_\xi$ of the form:
\begin{subequations}
\begin{align}
\hat{\cal H}_R &= -\frac{\hbar^2}{2M}\nabla_R^2 + \frac{M\Omega^2}{2}\boldsymbol{R}^2, \label{HamCM}\\
\hat{\cal H}_\xi &= -\frac{\hbar^2}{2\mu}\nabla_\xi^2 + \frac{\mu\Omega^2}{2}\boldsymbol{\xi}^2 + {\cal V}\left(|\boldsymbol{\xi}|\right), \label{HamRel}
\end{align}
\end{subequations}
where $M=2m$ and $\mu=m/2$. Consequently, every eigenstate of \eqref{Hamiltonian} can be written as $\Upsilon(\boldsymbol{r}_1,\boldsymbol{r}_2)=\Phi(\boldsymbol{R})\Psi(\boldsymbol{\xi})$. The center-of-mass Hamiltonian \eqref{HamCM} describes the single particle of mass $M$ confined in a two-dimensional harmonic trap and its eigenfunctions $\Phi_{NL}(\boldsymbol{R})$ with corresponding eigenenergies can be found straightforwardly. They are enumerated with two quantum numbers $(N,L)$ related to excitations in the radial direction and the angular momentum of the center of mass, respectively. The relative motion Hamiltonian \eqref{HamRel} is equivalent to the Hamiltonian of a single particle of mass $\mu$ confined in a two-dimensional harmonic trap imposed in the center by the additional rectangular potential ${\cal V}(|\boldsymbol{\xi}|)$. Our aim is to analyze all the properties of the Hamiltonian $\hat{\cal H}_\xi$. For convenience we express all quantities in the natural units of the harmonic oscillator, {\it i.e.}, energies, positions, and momenta are measured in $\hbar\Omega$, $\sqrt{\hbar/m\Omega}$, and $\sqrt{\hbar m\Omega}$, respectively.

Note that the Hamiltonian $\hat{\cal H}_\xi$ commutes with the relative angular momentum operator $\hat{\boldsymbol{L}}=-i\,\boldsymbol{\xi}\times\partial/\partial\boldsymbol{\xi}$. Therefore, to reduce complexity of the problem we rewrite it to the polar coordinates $\boldsymbol{\xi}=(\rho,\phi)$ and we represent all its eigen wave functions $\Psi(\boldsymbol{\xi})$ in the standard angular momentum representation
\begin{equation}
\Psi(\boldsymbol{\xi})=\Psi(\rho,\phi) = \frac{1}{\sqrt{\rho}}f(\rho)\mathrm{e}^{\pm i \ell \phi}.
\end{equation}
The wave function $\Psi(\rho,\phi)$ obeys the Schr\"odinger equation $(\hat{\cal H}_{\xi}-{\cal E})\Psi(\boldsymbol{\xi})=0$ iff $\ell=0, 1, 2,\ldots$ and the radial part $f(\rho)$ fulfills the one-dimensional radial Schr\"odinger equation of the form
\begin{align}\label{radial2}
\left[-\frac{\mathrm{d}^2}{\mathrm{d}\rho^2}+ \frac{\ell^2-\frac{1}{4}}{\rho^2}+\frac{\rho^2}{4}+{\cal V}(\rho)-{\cal E}\right]f(\rho)=0.
\end{align}
Due to the specific form of the interaction \eqref{interaction}, the eigenequation \eqref{radial2} has simplified form in the two disjoint regions $\rho<a$ and $\rho>a$:
\begin{align} \label{radial}
\left[-\frac{\mathrm{d}^2}{\mathrm{d}\rho^2}+ \frac{\ell^2-\frac{1}{4}}{\rho^2}+\frac{\rho^2}{4}-E\right] f(\rho)=0
\end{align}
with $E={\cal E}$ and $E={\cal E}-V$, respectively. It is a matter of fact that for any $E$ the equation \eqref{radial} has two independent solutions (for $V\neq 0$) which can be expressed in terms of confluent hypergeometric functions $\mathbf U$ and $_1{\mathbf F}_1$ as follows:
\begin{subequations}
\begin{align}
f_{1}(\rho)&=\rho^{\ell+\frac{1}{2}}\,\mathrm{e}^{-\frac{\rho^2}{4}}\,_1{\mathbf F}_1\left(\frac{-{E}+\ell+1}{2};\ell+1;\frac{\rho^2}{2}\right), \\
f_{2}(\rho)&=\rho^{\ell+\frac{1}{2}}\,\mathrm{e}^{-\frac{\rho^2}{4}}\,{\mathbf U}\left(\frac{-{E}+\ell+1}{2},\ell+1,\frac{\rho^2}{2}\right).
\end{align}
\end{subequations}
Since the function $f_1(\rho)$ is divergent in the limit $\rho\rightarrow\infty$, any physically acceptable solution of \eqref{radial} can be constructed only as the composition

\begin{equation} \label{ffunc}
f(\rho) \propto \left\{
\begin{tabular}{ll}
$A f_{1}(\rho)$, \qquad &$\rho<a$ \\
$ f_{2}(\rho)$, \qquad & $\rho\geq a$
\end{tabular}
\right.\,
\end{equation}
with appropriately chosen energy ${\cal E}$ and coefficient $A$ to match both parts at $\rho=a$ and to assure that the wave function $f(\rho)$ is continuous and differentiable in a whole space
\begin{align}\label{trans}
Af_{1}(a)-f_{2}(a) = 0 =
\left.\frac{\mathrm{d}}{\mathrm{d}\rho}\Big[Af_{1}(\rho)- f_{2}(\rho)\Big]\right|_{\rho=a}.
\end{align}
These conditions can be fulfilled only for appropriately chosen (quantized) energies ${\cal E}_{n\ell}$. After simple algebra one finds that the eigenenergies ${\cal E}_{n\ell}$ must be solutions of the following transcendental equation
\begin{multline} \label{energycond}
(-{\cal E}_{n\ell}+V+\ell+1)\,{\mathbf U}\left(\frac{-{\cal E}_{n\ell}+\ell+1}{2},\ell+1,\frac{a^2}{2}\right)\, _1{\mathbf F}_1\left(\frac{-{\cal E}_{n\ell}+V+\ell+3}{2};\ell+2;\frac{a^2}{2}\right) \\
-(\ell+1) ({\cal E}_{n\ell}-\ell-1)
   {\mathbf U}\left(\frac{-{\cal E}_{n\ell}+\ell+3}{2},\ell+2,\frac{a^2}{2}\right)\, \, _1{\mathbf F}_1\left(\frac{-{\cal E}_{n\ell}+V+\ell+1}{2};\ell+1;\frac{a^2}{2}\right)=0
\end{multline}
where the quantum number $n=0,1,\ldots$ enumerates successive roots of the equation \eqref{energycond}. Finally, the eigenstates of the relative motion Hamiltonian $\hat{\cal H}_{\xi}$ are determined by two quantum numbers $(n,\ell)$ and the angular momentum orientation and they have the form
\begin{equation}
\Psi^{(\pm)}_{n\ell}(\boldsymbol{\xi}) \propto \frac{1}{\sqrt{\rho}}f_{n\ell}(\rho)\mathrm{e}^{\pm i \ell \phi},
\end{equation}
where $f_{n\ell}(\rho)$ is given by \eqref{ffunc} provided that the eigenenergy ${\cal E}_{n\ell}$ is the $n$-th root of the transcendental equation \eqref{energycond}.
\begin{table}
\centering
\begin{tabular}{||clc||}
\hline \hline
\mbox{}\hspace{0.5cm} & $\bullet$ $\mathbf{V=4}$, $ \mathbf{a=2}$ &\\
 &\qquad $(n,\ell)=(0,0)$ and ${\cal E}_{00}=3$, &\\
 &\qquad $f_1(\rho) = \rho^{1/2}\,\mathrm{e}^{\,\rho^2/4}$, &\\
 &\qquad $f_2(\rho) =\rho^{1/2}\, \left(\frac{1}{2}\rho^2-1\right)\mathrm{e}^{-\rho^2/4}$, &\\
 &\qquad $A=\mathrm{e}^{-2}$. & \\
\hline
 &$\bullet$ $\mathbf{V=6}$, $\mathbf{a=\sqrt{6}}$ & \\
 &\qquad $(n,\ell)=(0,1)$ and ${\cal E}_{01}=4$, &\\
 &\qquad $f_1(\rho)=\rho^{3/2}\,\mathrm{e}^{\,\rho^2/4}$, &\\
 &\qquad $f_2(\rho)=\rho^{3/2}\left(\frac{1}{2}\rho^2-2\right)\mathrm{e}^{-\rho^2/4}$, &\\
 &\qquad $A=\mathrm{e}^{-3}$. &\\
\hline
 &$\bullet$ $\mathbf{V=6}$, $\mathbf{a=(6-2\sqrt{3})^{1/2}}$ &\\
 &\qquad $(n,\ell)=(1,0)$ and ${\cal E}_{10}=5$, &\\
 &\qquad $f_1(\rho)=\rho^{1/2}\,\mathrm{e}^{\,\rho^2/4}$, &\\
 &\qquad $f_2(\rho)=\rho^{1/2}\,\left(\frac{1}{4}\rho^4-2 \rho^2+2\right)\mathrm{e}^{-\rho^2/4}$, &\\
 &\qquad $A=-2(\sqrt{3}-1)\mathrm{e}^{\sqrt{3}-3}$. &\\
\hline
 &$\bullet$ $\mathbf{V=8}$, $\mathbf{a=2}$ &\\
 &\qquad $(n,\ell)=(1,1)$ and ${\cal E}_{11}=6$, &\\
 &\qquad $f_1(\rho)=\rho^{3/2}\,\mathrm{e}^{\,\rho^2/4}$, &\\
 &\qquad $f_2(\rho)=\rho^{3/2}\,\left(\frac{1}{4}\rho^4-3\rho^2+6\right)\mathrm{e}^{-\rho^2/4}$, &\mbox{}\hspace{0.5cm} \\
 &\qquad $A=-2\mathrm{e}^{-2}$. & \\
\hline
\hline
\end{tabular}
\caption{Several examples of potential parameters $V$ and $a$ for which the solutions of the eigenproblem \eqref{radial2} have simplified algebraic form. These particular solutions may serve as benchmarks for different numerical approaches. \label{_Table1}}
\end{table}

At this point, it is interesting to point out that for some particular potential parameters $V$ and $a$ the exact solutions may be significantly simplified. It may happen when the first arguments of the confluent hypergeometric functions $\mathbf{U}$ and $_1\mathbf{F}_1$ are integers and the functions are expressed in terms of simple algebraic expressions \cite{abramowitz1965}. For example, when $V=4$ and $a=2$ one finds the ground-state energy ${\cal E}_{00}=3$ and consequently $f_1(\rho)=\rho^{1/2}\,\mathrm{e}^{\rho^2/4}$, $f_2(\rho)=\sqrt{\rho}(\rho^2/2-1)\mathrm{e}^{-\rho^2/4}$, and $A=\mathrm{e}^{-2}$. We present some other examples in Table~\ref{_Table1}. Such simple algebraic solutions play a very important role, since they may serve as benchmarks for the accuracy of different numerical techniques.

Finally, we want to emphasize that up to now the solutions are in fact obtained for distinguishable particles described by the Hamiltonian \eqref{Hamiltonian}. In the case of indistinguishable atoms (fermions or bosons), one should impose additional requirements to the wave functions of the relative motion under exchange of particles' positions, $\boldsymbol{\xi}\rightarrow-\boldsymbol{\xi}$. In consequence, the bosonic (fermionic) states have even (odd) angular momentum quantum numbers~$\ell$. It means that the radial distribution of fermionic relative motion must necessarily vanish at $\rho=0$. This fact can be viewed as a direct manifestation of the Pauli exclusion principle forbidding any two identical fermions to occupy the same position.

\section{Eigenstates classification} \label{Sec:Porownanie}
\begin{figure}
\centering
\includegraphics[scale=0.9]{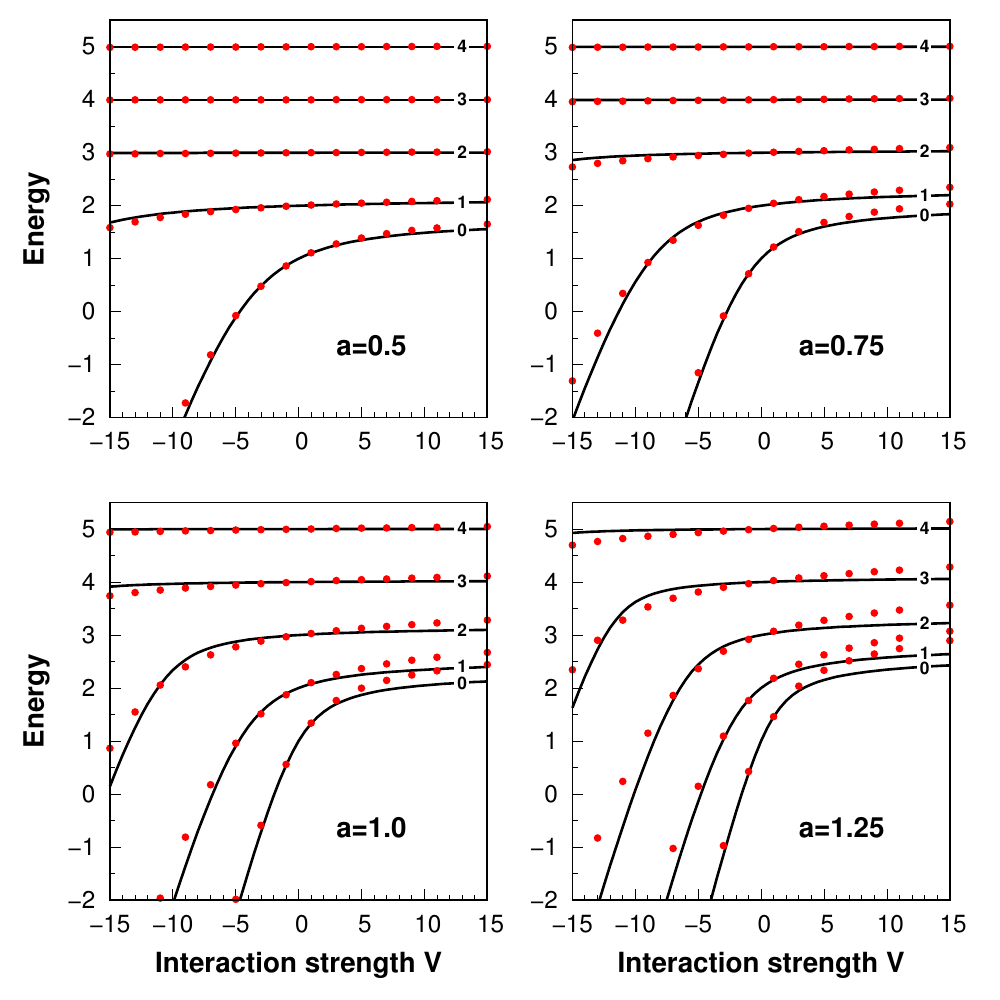}
\caption{The ground-state energy of the relative motion Hamiltonian \eqref{HamRel} in the subspaces of given relative angular momentum $\ell=0,\ldots,4$ (labels on solid lines) for different values of the potential range $a$. Solid lines are obtained for simplified potential ${\cal V}$ while red dots for realistic interaction potential $\cal V_R$ with $R_c=3a/\pi$. It is clearly seen that for ranges smaller than the natural length of the harmonic oscillator ($a\lessapprox 1$) both approaches give very similar results in a wide range of interaction strengths. \label{_Fig2}}
\end{figure}
It is very instructive to start the analysis of the system's properties focusing on the ground-state energy of the relative motion Hamiltonian \eqref{HamRel} in individual subspaces of given relative angular momentum, {\it i.e.}, states with $(n,\ell)=(0,\ell)$. In Fig.~\ref{_Fig2} we present resulting spectra for several potential ranges $a\in\{0.5, 0.75, 1.0, 1.25\}$. For completeness, we compare eigenenergies obtained in our model of interaction potential ${\cal V}$ (solid black lines) with those obtained numerically when the realistic model of mutual interactions ${\cal V}_R$ is considered (red dots). It is clear that for not too large ranges of the potential, predictions of both approaches agree in a wide range of interaction strengths. Deviations are clearly visible for large potential ranges $a\gtrapprox 1$ and/or adequately strong interactions $V\gtrapprox 7$.

\begin{figure}
\centering
\includegraphics[scale=0.9]{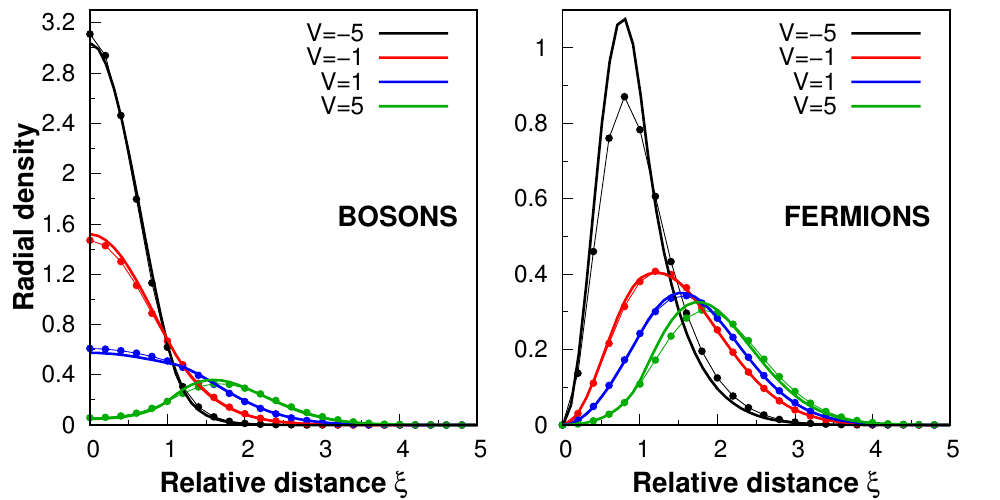}
\caption{Radial density distribution $F(\rho)$ of the relative motion in bosonic $(n,\ell)=(0,0)$ and fermionic $(n,\ell)=(0,1)$ ground states for $a=1$ and different strengths of interaction $V$. Thick solid lines represent distributions obtained in the simplified model of interactions \eqref{interaction} while thin dotted lines represent results in the realistic model \eqref{RydbergPot}. Note that in the case of fermionic particles, due to the Pauli exclusion principle, the radial density distribution $F(\rho)$ must necessarily vanish at $\rho=0$. \label{_Fig3}}
\end{figure}

Note, that even for quite large potential range, $a=0.5$, and quite large strengths $V$, only the $s$-state with $\ell=0$ is influenced by interactions. It is clear that states which are the most sensitive to interactions are characterized by the smallest relative angular momentum quantum numbers $\ell$. Along with increasing $\ell$, radial distributions of relative motion are pushed out from the center due to the additional centrifugal term in the Hamiltonian. In consequence, they are less sensitive to the interaction core. This effect is clearly seen when the radial density distribution of the relative motion is considered, $F(\rho) = f^2(\rho)/\rho$. In Fig.~\ref{_Fig3} we plot this distribution for the bosonic $(n,\ell)=(0,0)$ and the fermionic $(n,\ell)=(0,1)$ ground states of the relative Hamiltonian \eqref{HamRel} and different potential strengths $V$ (here we set $a=1$). For increasing repulsions the density probability is suppressed in the center, while for attractions it is enhanced. In fact, this effect almost does not depend on long-range tails of the potential and it is an exclusive consequence of the potential core. It is clear when we compare the distributions obtained in the simplified model of interaction ${\cal V}$ (solid lines) with predictions of the realistic model ${\cal V}_R$ (dotted lines). In both cases, the resulting distributions are almost identical (see Fig.~\ref{_Fig3}). In the limiting case, $a\rightarrow 0$, only the $s$-states ($\ell=0$) are affected by interactions since only these states have non-vanishing distributions at $\rho=0$.

\begin{figure}
\centering
\includegraphics[scale=0.9]{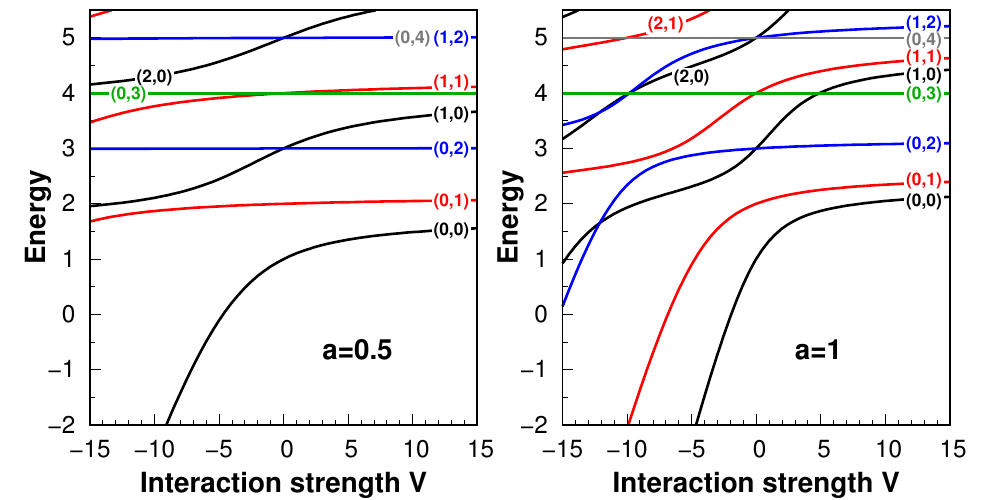}
\caption{Spectrum of the relative motion Hamiltonian \eqref{HamRel} as a function of the interaction strength $V$ presented for two representative values of the potential range $a=0.5$ and $a=1.0$. Different colors correspond to states with different relative angular momentum $\ell$. For clarity all lines are labeled with their quantum numbers $(n,\ell)$. \label{_Fig4}}
\end{figure}

The situation becomes even more interesting when excitations of the relative motion are considered. In Fig.~\ref{_Fig4} we present the energy spectrum of the relative motion Hamiltonian \eqref{HamRel} for two different potential ranges classified accordingly with their quantum numbers $(n,\ell)$. As it is seen, for interactions having larger potential ranges, an energetic order of eigenstates can be changed. Moreover, almost perfect degeneracies between different states visible for smaller ranges are lifted (compare behavior of states $\{(0,4),(1,2)\}$ or $\{(1,1),(0,3)\}$ for $a=0.5$ and $a=1.0$). Finally, let us draw some attention to the effect of decreasing splitting between the two lowest eigenstates $(0,0)$ and $(0,1)$ (the latter is in fact doubly degenerated due to the orientation of the angular momentum) when larger ranges are considered. Vanishing of this particular gap may have crucial experimental consequences for distinguishable particles since then even very small but a finite temperature of the system may lead to the statistical mixing of these states and significantly change measurable properties of the system. On the other hand, in the situation of a small gap the states can be easily coupled by some additional well-controlled but non-conserving angular momentum interactions  (for example spin-orbit coupling).

\section{Inter-particle correlations} \label{Sec:Cor}
\begin{figure}[h!]
\includegraphics[scale=0.9]{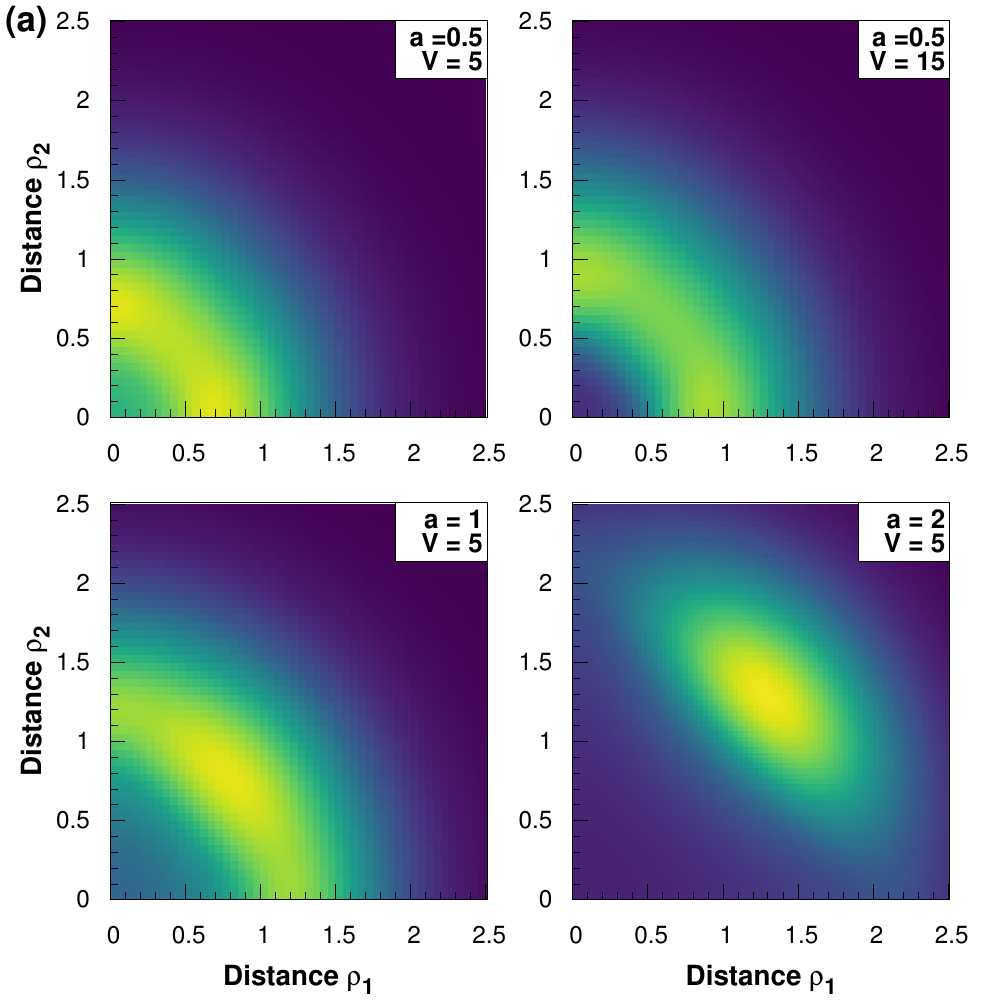} 
\includegraphics[scale=0.9]{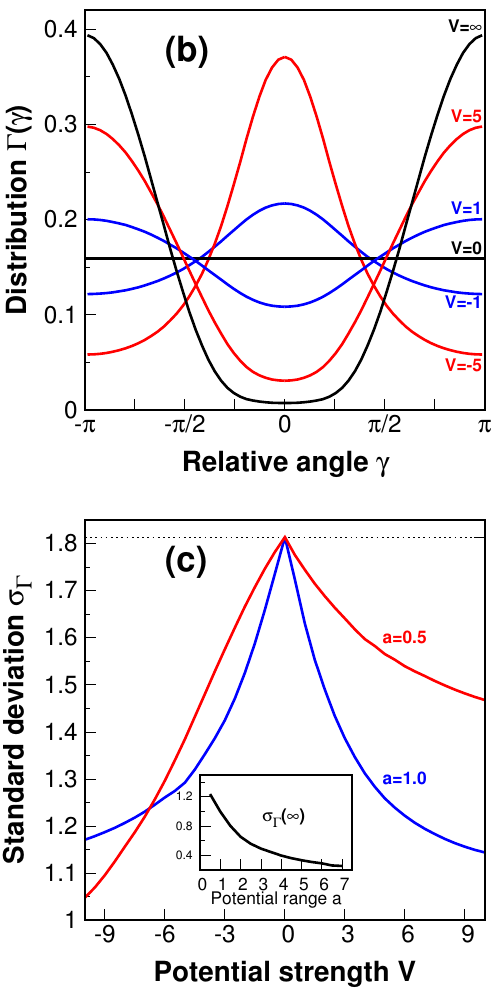}
\caption{Inter-particle correlations in the bosonic ground-state $(N,L,n,\ell)=(0,0,0,0)$ of interacting Rydberg-dressed atoms. (a) The two-particle radial distribution $n(\rho_1,\rho_2)$ for four representative sets of parameters characterizing the interaction potential. (b) The two-particle azimuthal distribution $\Gamma(\gamma)$ for $a=1$ and different values of interaction strength $V$. (c) The standard deviation $\sigma_\Gamma(V)$ of the distribution $\Gamma(\gamma)$ as a function of potential strength $V$ for two different values of potential range $a$. Maximal value of the standard deviation reached at $V=0$ is equal to $\pi/\sqrt{3}$ and corresponds to the flat distribution $\Gamma(\gamma)=(2\pi)^{-1}$ [horizontal line in plot (b)]. The inset displays values of the standard deviation $\sigma_\Gamma$ reached for an infinite potential strength ($V\rightarrow\infty$) as a function of the potential range~$a$. \label{_Fig5}}
\end{figure}
Having analytical expressions for two-particle eigenstates of the interacting system $\Upsilon_{NL;n\ell}(\boldsymbol{R},\boldsymbol{\xi})=\Phi_{NL}(\boldsymbol{R})\Psi_{n\ell}(\boldsymbol{\xi})$ it is very easy to perform the inverse transformation and obtain two-particle wave functions $\Upsilon_{NL;n\ell}(\rho_1,\varphi_1;\rho_2,\varphi_2)$ expressed by particles' real-space positions $\boldsymbol{r}_1=(\rho_1,\varphi_1)$ and $\boldsymbol{r}_2=(\rho_2,\varphi_2)$. Then, one can straightforwardly analyze different interesting features of inter-particle correlations. Here we focus on the two simplest quantities which directly encode information about {\it relative spatial correlations} between particles. The first is the two-particle radial distribution $n(\rho_1,\rho_2)$ defined as
\begin{equation}
n(\rho_1,\rho_2) = \int\!\!\mathrm{d}\varphi_1\,\mathrm{d}\varphi_2 \left|\Upsilon(\rho_1,\varphi_1;\rho_2,\varphi_2)\right|^2, \end{equation}
which can be directly interpreted as the probability density that simultaneously observed particles are at distances $\rho_1$ and $\rho_2$ from the center of the trap (see Fig.~\ref{_Fig5}a). It is evident that along with increasing interaction strength $V$ particles are pushed out from the center of the trap and their radial distances become correlated, {\it i.e.}, the probability of finding particles at the same distance from the center increases. The second quantity is the two-particle azimuthal distribution $\Gamma(\varphi_1,\varphi_2)$ defined as
\begin{equation} \label{GammaDef}
\Gamma(\varphi_1,\varphi_2) = \int\!\!\rho_1\rho_2\,\mathrm{d}\rho_1\,\mathrm{d}\rho_2 \left|\Upsilon(\rho_1,\varphi_1;\rho_2,\varphi_2)\right|^2.
\end{equation}
It is related to the probability that in a simultaneous measurement of particles' positions the position vectors will be oriented at angles $\varphi_1$ and $\varphi_2$, respectively. If the two-particle quantum state is rotationally invariant (for example the two-boson ground-state of the system has this property) then the distribution \eqref{GammaDef} depends only on a difference $\gamma=\varphi_1-\varphi_2$ and then one can consider the simplified distribution $\Gamma(\gamma)=2\pi\Gamma(\gamma+\varphi_0,\varphi_0) $ (with arbitrary chosen $\varphi_0$) encoding probability density for the relative angle between position vectors.
We plot the distribution $\Gamma(\gamma)$ for different interaction strengths $V$ and $a=1$ in Fig.~\ref{_Fig5}b. Obviously, for $V=0$ the distribution is flat and equal to $(2\pi)^{-1}$. When repulsive interactions are switched on, the probability that particles occupy opposed sides of the trap  ($\gamma=\pm \pi$) is strongly enhanced. Contrary, for attractive interactions ($V<0$), particles are more likely to be located on the same side of the trap ($\gamma=0$). One can quantify an uncertainty that particles are exactly on the opposite (same) side by calculating the standard deviation defined as
\begin{equation}
\sigma_\Gamma^2 = 2\int_0^\pi \mathrm{d}\gamma\,\left(\overline{\gamma}-\gamma\right)^2\Gamma(\gamma)
\end{equation}
with $\overline{\gamma}=\pi$ and $\overline{\gamma}=0$ for repulsive and attractive interactions, respectively. We display this quantity in Fig.~\ref{_Fig5}c. It is clear that along with increasing interactions uncertainty decreases. Surprisingly, it decreases also when the range of the potential $a$ increases. It means that larger ranges try to force particles to form a line with the center of the trap. This observation is also supported by results obtained for infinite potential strength $V\rightarrow\infty$ (inset in Fig.~\ref{_Fig5}c). As it is seen, in this limit the standard deviation $\sigma_\Gamma(\infty)$ decreases with the potential range $a$. Interestingly, the effect of the potential range is not so obvious for attractive interactions. In this case, the behavior of the system strongly depends on potential strength. All these three quantities together [$n(\rho_1,\rho_2)$, $\Gamma(\varphi_1,\varphi_2)$, and $\sigma_\Gamma(V)$] give a quite nice view on the spatial correlations induced by interactions build in the system. It can be summarized as follows. When repulsions are increased, positions of particles are forced to arrange exactly on opposite sides of the trap in a quite well-established distance from the center which is determined by the potential range. The effect is stronger for larger $a$. Contrary, when interactions are attractive, the most probable situation is that particles are found on the same side of the trap.

\section{Generalization to three dimensions} \label{Sec:Gen}
Finally, let us also mention that the presented solutions may be easily generalized to the problem of two-particles confined in the isotropic three-dimensional harmonic trap. In this case, all eigenfunctions of corresponding three-dimensional relative motion Hamiltonian are classified by three quantum numbers $(n,\ell,m)$ and have a form
\begin{equation}
\Psi(\boldsymbol{\xi})=\Psi_{n\ell m}(\rho,\theta,\phi) = \frac{1}{\rho}f_{n\ell}(\rho)\mathrm{Y}_{\ell m}(\theta,\phi),
\end{equation}
where $\mathrm{Y}_{\ell m}(\theta,\phi)$ are three-dimensional spherical harmonic functions. The radial part of the eigenfunction $f_{n\ell}(\rho)$ fulfills the following single-particle Schr\"odinger equation
\begin{equation}
\left[-\frac{\mathrm{d}^2}{\mathrm{d}\rho^2}+ \frac{\ell(\ell+1)}{\rho^2}+\frac{\rho^2}{4}+{\cal V}(\rho)-{\cal E}_{n\ell}\right]f_{n\ell}(\rho)=0.
\end{equation}
This eigenproblem is exactly equivalent to the previous two-dimensional problem \eqref{radial2} provided that one perform appropriate substitution $\ell \mapsto \ell + 1/2$ in \eqref{radial2}. Consequently, by applying this substitution in \eqref{ffunc} and \eqref{energycond}, one obtains three-dimensional eigenstates and transcendental equation for eigenenergies, respectively.

\section{Summary} \label{Sec:Conclu}
To conclude, in our work we introduced a simplified model of two interacting Rydberg-dressed atoms confined in a harmonic trap. The model is a consequence of replacing the realistic shape of interaction potential by the soft-core finite-range forces modeled by a step function. The main advantage of the model proposed is its exact solvability in terms of special functions. This gives a route for analytical analysis of different properties of the system which are crucially important when inter-particle correlation are considered. By performing detailed numerical analysis, we show that the eigenstates obtained in the simplified model are very close to those obtained in a realistic model in a wide range of interaction parameters. Although our work is devoted only to two interacting atoms, the solutions presented can be used as building-blocks for approximate methods dedicated to a larger number of particles. For example, similarly as it was done in different one-\cite{2012CremonFBS,2012BrouzosPRL,2013BrouzosPRA}, two-\cite{2011BertainaPRL,2017LutsyshynJCP}, or three-dimensional\cite{1977KrotscheckPRA}, one of the possible extensions is to use these two-particle solutions when variational ansatz of {\it pair-correlated} Jastrow wave functions are constructed \cite{1955JastrowPhysRev}. Typically for this construction of the variational family, the corresponding two-body exact solution may serve as a prescription for trial wave functions.
 
Finally, we want to point out that in cases of more than two interacting Rydberg atoms, the simplified model of interactions should be used carefully since neglecting the long-range part of interactions may lead to false conclusions. For sure, the simplified soft-core finite-range potential captures the most important part of realistic interactions between Rydberg atoms. As long as long-range tails do not affect (or affect much weaker) a third particle being far from considered pair, the simplified model should appropriately describe properties of the many-body system. It means that the simplified model of interactions gives an appropriate description only when two-body interactions are not substantially affected by long-range interactions with other particles. In other cases, the long-range tails may substantially change properties of the system and introduce additional correlations.

\bibliography{biblio.bib}

\section*{Acknowledgements}
The authors would like to thank Jacek Dobrzyniecki for his comments and suggestions. This work was supported by the (Polish) National Science Center Grant No. 2016/22/E/ST2/00555 (TS).

\section*{Author contributions statement}
P.K. and T.S. equally contributed in all stages of the project.

\section*{Additional information}

\textbf{Competing financial and non-financial interests} Both authors declare no competing interests.

\end{document}